\documentclass[12pt]{article}
\usepackage{amsmath}
\usepackage{amssymb}
\usepackage{graphicx}
\usepackage[cp1251]{inputenc}
\usepackage[russian]{babel}
\newcommand{\be}{\begin{equation}}
\newcommand{\ee}{\end{equation}}
\newcommand{\ba}{\begin{eqnarray}}
\newcommand{\ea}{\end{eqnarray}}

\begin{document}
\vspace{1cm}

\begin{center}
\bf RKKY interaction of magnetic moments in nanosized  systems   \\
\end{center}
\bigskip

\centerline{E.Z Meilikhov, and R.M. Farzetdinova}
\medskip
\centerline{\small\it Russian research center “Kurchatov Institute”,
123182 Moscow, Russia}
  \vspace{1cm}

  \centerline{
\begin{tabular}{p{15cm}}
\footnotesize \hspace{15pt} Nanosized spherical system of magnetic
moments interacting indirectly via the RKKY mechanism is studied.
The interaction energy that determines the temperature of the
ferromagnetic ordering, depends strongly on the system size.
Obtained in the mean-field approximation, dimensional and
concentration dependencies of the Curie temperature testify to the
necessity of taking into account the finite size of such systems to
calculate their features. Results may concern both artificially
constructed nanosystems and naturally arising formations (such as
clusters of magnetic ions in diluted magnetic semiconductors, etc.).
\end{tabular}
} \vspace{1cm}

In systems with free carriers of high concentration (metals or
degenerate semiconductors), indirect magnetic impurities'
interaction of Ruderman-Kittel-Kasuya-Yosida (RKKY) type is
considered as one of the basic mechanisms of the magnetic
ordering~\cite{1}. There are a lot of papers dealing with RKKY
interaction in three-, two- and one-dimensional systems of
\emph{infinite} size~\cite{1,2,3,4}. However, to our knowledge,
nobody considered how that interaction should be modified for
systems of the \emph{finite} size. In the present paper, we consider
this problem for the case of the spherical system of the finite
radius.

Estimating the energy $w(r)=-J(r){\bf S}_1{\bf S}_2$ of the indirect
RKKY-interaction of magnetic impurities with spins  ${\bf S}_1$,
${\bf S}_2$  spaced at the distance ${\bf r}$  is based on making
use of the expression
 \be\label{1}
J(r)=\frac{J_{pd}^2}{N^2}\exp(-r/l){\sum\limits_{\bf
q}}'\sum\limits_{\bf k} e^{i{\bf qr}}\frac{f(E_{\bf k})}{E_{\bf
k+q}-E_{\bf k}},
 \ee
obtained in the second order of the perturbation theory~\cite{1}.
Here, $N$ is the number of lattice sites, $J_{pd}$ is the exchange
energy for the interaction of the impurity spin with a free charge
carrier, $E_{\bf k}$ is the carrier energy, $f(E_{\bf k})$ is the
Fermi-Dirac function which in the degenerate case equals $f(E_{\bf
k})=1$ at $k<k_F$ and $f(E_{\bf k})=0$ at $k<k_F$, where $k_F$ is
the Fermi momentum. The prime by the the sum over $q$ means that
$q\ne0$. The exponent $e^{-r/l}$ in (\ref{1}) reflects the finite
carrier mean free path $l$.

In the continual approximation, the summation in (\ref{1}) is
replaced by the integration  which performed usually over \emph{all}
$k<k_F$ and {$|\bf{k}+\bf{q}|>k_F$}. Then, the standard result
corresponding to the case of the infinite system reads~\cite{1}:
 \be\label{2}
J(r)=-I_0 \Phi(r)\exp(-r/l),\quad
\Phi(r)=\left(\frac{a}{r}\right)^4[\varphi(r)\cos\varphi(r)-\sin\varphi(r)],
 \ee
where
 \be\label{3}
I_0=\frac{1}{32\pi^3}\left(\frac{ma^2}{\hbar^2}J_{pd}^2\right),\quad
\varphi(r)=2k_{F0}r,
 \ee
$a$ is the lattice constant, $k_{F0}=(3\pi^2p)^{1/3}$  is the Fermi
momentum of carriers of the concentration~$p$.

In the case of finite system sizes or magnetic ions' clustering, the
classic expression~(\ref{2}) for the energy of the RKKY-interaction
should be rectified. For simplicity, we consider the case when
magnetic ions form the spherical cluster. Due to the
quasi-neutrality,  its radius $R$ determines not only the area where
ions are arranged but also the region where carriers, produced by
those ions, are localized. In other words, the carriers  are
contained in the potential well of the radius~ $R$. Therefore, the
carrier momentum $k$ and its variation $q$ are limited by the
intervals
 \be\label{4}
k_1\le k\le k_F,\quad k_1\le q\le k_2,
 \ee
where
\be\label{xi}
 k_1\approx\pi/ R,\,k_2\approx\pi/a.
 \ee
In addition, due to the spatial quantization the distance between
energy levels of carriers grows that leads to increasing the Fermi
energy and Fermi momentum with decreasing the well size:
$k_F=k_F(R)$. Together, it  complicates calculations and the final
expression for $J(r)$ turns out to be more bulky than the canonical
expression~(\ref{2}).

The finite mean free path $l$ results in smearing energy levels of
carriers due to their collisions. Therefore, the lowest value $k$ is
defined by the system size and equals $k_1\approx\pi/R$ only if the
collision broadening $\hbar/\tau$  of levels is less then the energy
$\pi^2\hbar^2/2mR^2$ of the first level. That condition could be
written in the form
 \be\label{condition}
\frac{\pi^2}{R^2}-\frac{2k_F}{l}>0
 \ee
meaning that our approach relates to the small enough systems only.
If, for instance, $l/a=10$, $ak_F=1$ then $R\lesssim10a$.  In the
general case,  one could use the value
 \be\label{k1}
k_1=\sqrt{
    {\rm max}\left\{\left(\frac{\pi^2}{R^2}-\frac{2k_F}{l}\right),\,0\right\}
    }
 \ee
as the left boundary of inequalities (\ref{4}).

To proceed one should estimate how the Fermi momentum depends on the
cluster size. The total number of  free carriers in the cube of the
size $R$ with the spherical Fermi surface is defined by the number
of cells of the volume $(2\pi)^3$ in the phase space and in the
limit of $k_FR\to\infty$ equals $N_F\sim k_F^3R^3/(2\pi)^3$. For a
finite $k_FR$-value, the number of carriers is defined by the number
of points in the wavenumber space with coordinates divisible by
$(2\pi/R)$. As none of those coordinates could  equal zero, in that
case
$$
N_F\sim (k_FR/2\pi)^3 -\alpha(k_FR/2\pi)^2,
$$
where the correction (proportional to $\alpha\sim1$) is associated
with the “forbidden” points, positioned in the coordinate planes of
the wavenumber space and with the “correct” points neighboring to
the Fermi surface but not fallen inwards, as well. For the carrier
concentration $n=N_F/R^3$, it follows $n\sim k_F^3(1-\alpha/k_FR)$.
For $n={\rm const}$ and $k_FR\gg1$ that leads to
$$
k_F(R)- k_{F0}\approx\frac{\alpha}{3 R},
$$
where $k_{F0}\equiv k_F(\infty)$ is the Fermi momentum in the
infinitely large system. Everywhere below, we use $k_F$ to mean the
value
 \be\label{5}
k_F(R)=k_{F0}+\frac{1}{R}.
 \ee
\noindent Let us turn now to calculating the energy of the RKKY
interaction.  Assuming $E_{\bf k}=\hbar^2k^2/2m$, $E_{\bf
k+q}=\hbar^2(k+q)^2/2m$ and designating the angle between ${\bf r}$
and ${\bf q}$ as $\alpha$, and the angle between ${\bf k}$ and ${\bf
q}$ as $\theta$, we obtain
 \be\label{6}
J(r)=\frac{2mJ_{pd}^2}{\hbar^2N^2}\,\exp(-r/l)\,a^4\,{\sum\limits_q}'\frac{\,e^{i
qr\cos\alpha}}{q}\,\,\sum\limits_k \frac{f(E_k)}{2k\cos\theta+q},
 \ee
or, in the continual approximation,
 \be\label{7}
J(r)=\frac{4}{\pi}\,I_0\exp(-r/l)\,a^4\!\!\int\limits_{k_1}^{k_2}qdq
\int\limits_0^\pi e^{i qr\cos\alpha}\sin\alpha d\alpha
 \int\limits_{k_1}^{k_{F}}k^2dk
 \int\limits_0^\pi\frac{\sin\theta d\theta}{2k\cos\theta+q}.
 \ee

Non-complicated but laborious calculations lead to rather cumbersome
result which could be represented in the relatively simple form by
the help of the operator $\hat{L}$ determining the value of the
double definite integral
$\int_{k_1}^{k_{F}}dk\int_{k_1}^{k_2}(...)dq$ with the primitive
function $\Psi(k,q)$:
 $$
{\hat
L}\Psi(k,q)=\Psi(k_1,k_1)-\Psi(k_{F},k_1)+\Psi(k_{F},k_2)-\Psi(k_1,k_2).
$$
Then
 \be\label{8}
J(r)=\frac{1}{\pi}\,I_0\exp(-r/l)\,(ak_F)^4\hat{L}\Psi_r(k,q),
 \ee
where
 \ba\label{24}
 \Psi_r(k,q)=\frac{1}{(2k_Fr)^4}\left[\Phi_{\rm
1}(k)\,c_1(k,q)-\Phi_2(k)\,c_2(k,q)+\Phi_3(k,q)\right],\mbox{\hspace{70pt}}\\\nonumber\\
\Phi_{\rm 1}(k)=2kr\cos 2kr-\sin2kr ,\quad\Phi_2(k)=2kr\sin 2kr+\cos
2kr,\mbox{\hspace{50pt}}\\\nonumber\\ c_1(k,q)= {\rm
Si}[(2k+q)r]-{\rm Si} [(2k-q)r],\quad
c_2(k,q)={\rm Ci}[(2k+q)r]-{\rm Ci} (|2k-q|r),\\\nonumber\\
\Phi_3(k,q)=\cos qr(1+2k^2r^2)\ln\left|\frac{2k+q}{2k-q}\right|+qr (
 \sin qr-\frac{1}{2}qr\cos qr )
\left(\ln\left|\frac{2k+q}{2k-q}\right|-\frac{4k}{q}\right),\\\nonumber\\
{\rm Si}(x)=\int\limits_0^x\frac{\sin t}{t}\,dt, \quad {\rm
Ci}(x)=-\int\limits_{x}^\infty\frac{\cos
t}{t}\,dt.\mbox{\hspace{100pt}}\\\nonumber
 \ea
To the traditional situation ($R\to\infty$) there correspond $k_1\to
0$, $k_2\to\infty$. In that case $c_1\to\pi$, $c_2\to0$,
$\Phi_3\to0$, and
$${\hat L}\Psi\to\frac{\pi}{(2k_Fr)^4}\Phi_{\rm
1}(k_F)=\pi\frac{2k_Fr\cos 2k_Fr-\sin2k_Fr}{(2k_Fr)^4}.
$$
Hence, (\ref{8}) reduces to the standard expression (\ref{2}). For
finite values $k_1$, $k_2$, the interaction energy $J(r)$ should be
calculated with Eqs. (\ref{8})--(15).

The local effective RKKY-field $H_{\rm RKKY}$, generated in a given
point, is defined by the relation $\mu H_{\rm RKKY}=\sum_i J(r_i)$
where $r_i$ is the distance from that point to the $i$th magnetic
impurity, $\mu$ is the impurity magnetic moment. In the continual
approximation, the sum could be replaced by the integral $\mu H_{\rm
RKKY}=\int J(r')d^3r'$ where the integration is spread over the
volume occupied by impurities. Contrary to the infinite system, the
value of that integral depends on the position of the considered
point. For the spherical system, the effective field could be
characterized by the value
 \be\label{17}
 \mu H^0_{\rm RKKY}= 4\pi\!\!\int\limits_{r_{\rm min}}^R J(r)  r^2dr
 \ee
of that integral in the center of the sphere. Here $r_{\rm min}$ is
the minimum distance between impurities determined by the
discreteness of the crystal lattice (for instance, the minimum
distance between extrinsic Mn atoms, replacing Ga atoms in GaAs
lattice, amounts  $r_{\rm min}=a/\sqrt{2}\approx 4$\,\AA).

How much would the local field $H_{\rm RKKY}$ be non-uniform inside
the sphere one could judge by making note that in the case with the
interaction length  $l$ being comparable or shorter than the radius
$R$, the local field at the surface of the sphere should be
approximately half as large as its value in the center of the
sphere. It is not hard to show that for any point inside the sphere
being offset by the distance $h\le R$ from its center,  one could
employ the relation
 \be\label{18}
 \mu H_{\rm RKKY}= 4\pi\!\!\int\limits_{r_{\rm min}}^{R+h}J(r)
 r^2F(r)dr,\quad F(r)=\left\{\begin{tabular}{ll}
                                         1, &$r< R-h$\\
                                        $\frac{\displaystyle R^2-(h-r)^2}{\displaystyle 4rh}$,&$ R-h<r<R+h$\\
                                         \end{tabular},\right.
 \ee
instead of (\ref{17}). In particular, for the field at the sphere
surface ($h=R$) one could find
 \be\label{19}
 \mu H^S_{\rm RKKY}= 2\pi\!\!\int\limits_{r_{\rm min}}^{2R}
 J(r)(1-r/2R)r^2dr.
 \ee

The results of numerical calculations (see Fig.~1) with $r_{\rm
min}=a/\!\sqrt{2}$, $l=3a$, $k_{F0}a=1$ show  that for $R=10a$   the
part of the sphere where the effective field differs from $H^0_{\rm
RKKY}$ no more than by 20\%, amounts about 85\% of its volume, and
the average field $\bar H_{\rm RKKY}=(3/R^3)\int_0^R H_{\rm
RKKY}(h)h^2dh\approx0.93H^0_{\rm RKKY}$ (the same for $R=3a$ amounts
$\bar H_{\rm RKKY}\approx1.14H^0_{\rm RKKY}$). Hence, one could, in
the first approximation, ignore the non-uniformity of the effective
field and consider it as nearly uniform and equal to $\bar H_{\rm
RKKY}\approx H^0_{\rm RKKY}$.

In the mean-field theory, Curie temperature $T_{\rm C}$ of the
ferromagnetic state arising due to the RKKY-interaction is defined
by the simple relation~\cite{1}
 \be\label{20}
k_BT_{\rm C}\sim\mu\bar H_{\rm RKKY}\approx\mu H^0_{\rm RKKY}.
 \ee
It is of interest to understand how the so-defined  Curie
temperature depends on the system size $R$ at $k_{F0}={\rm const}$
or  varies with the carrier concentration (determined by the Fermi
momentum~\cite{5}) in systems of  fixed (but different)
sizes~\cite{6}. Corresponding dependencies are shown in Fig. 2, 3.

The dependence $T_{\rm C}(R)$ turns out to be non-monotone, with
pronounced oscillations of the period $2a$ at higher $k_{F0}$
values. They are resulted from that part of the interaction energy
$J(r)$ which is defined by the function $\Phi_3(k,q)$ given by (15),
namely, by the terms $\sin qr$ and $\cos qr$ with $q=k_2=\pi/a$.
That provides for the observed period.

At $R\sim l$, the Curie temperature exceeds noticeably its value for
$R\to\infty$ (shown by dot lines in Fig. 2). But the most
significant distinction of the dependence $T_{\rm C}(R)$ (associated
with accounting the finite $k_1$ value) is observed at small $R$. As
soon as the condition (\ref{condition}) is reached, the drop of
$T_{\rm C}$ is observed, as compared to the “standard”
(corresponding to $k_1=0$) dependence (shown by dash curves in Fig.
2). That means the impossibility of existing ferromagnetism due to
RKKY interaction in small clusters.

As for the dependence $T_{\rm C}(k_{F0})$, the Curie temperature
increases almost monotonously with  $k_{F0}$ (i.e., with rising the
carrier concentration). In that case also, reaching the condition
(\ref{condition}) results in the considerable decreasing $T_{\rm C}$
(down to zero at some finite size $R\sim l$) as compared to the
$T_{\rm C}(k_{F0})$ dependence for $R\to\infty$.

In conclusion, we studied nanosized spherical systems of magnetic
moments interacting indirectly via the RKKY mechanism. The
interaction energy which determines the temperature $T_{\rm C}$ of
ferromagnetic ordering, depends strongly on the system size.
Obtained in the mean-field approximation, dimensional and
concentration dependencies of the Curie temperature testify to the
necessity of taking into account the finite size of such systems to
calculate their features. Results may concern both artificially
constructed nanosystems and naturally arising formations (such as
clusters of magnetic ions in diluted magnetic
semiconductors~\cite{7,8}, etc.).

This work has been supported by Grant No. 06-02-116313 of the
Russian Foundation of Basic Researches.

\newpage
\renewcommand{\refname}{\centerline{\rm \small\bf References}\vspace{5mm}}

\newpage
\centerline{\bf Captions}
\bigskip
\bigskip

Fig. 1. Spatial distribution of the local effective field $H_{\rm
RKKY}$ within the sphere of the radius $R=10a$ and $R=3a$ at
$l=3a$, $k_{F0}a=1$.\\

Fig.2. Dependencies $T_{\rm C}(R)$ of the Curie temperature on the
radius of  spherical systems with various carrier concentrations.
Upper panel: $l=3a$, lower panel: $l=10a$. Dot lines indicate the
values $T_{\rm C}(R=\infty)$, dash curves show the “standard”
(corresponding to $k_1=0$) behavior of the dependencies $T_{\rm
C}(R)$ at small $R$.\\

Fig. 3. Dependencies $T_{\rm C}(k_{F0})$ of the Curie temperature on
the carrier Fermi momentum  for spherical systems of various radii.
Upper panel: $l=3a$, lower panel: $l=10a$. Arrows indicate where the
condition~(\ref{condition}) is reached.

\newpage


\begin{thebibliography}{30}
\bibitem{1} D.C. Matiss, \emph{The theory of magnetism} (Harper
\& Row Publ., New York, Evanston, London, 1965).
\bibitem{2} Y. Yafet, Phys. Rev. B. {\bf 36}, 3948 (1987) --
1D-system.
\bibitem{3} B. Fisher and M. Klein, Phys. Rev. B. {\bf 11}, 2025 (1975) --
2D-system.
\bibitem{4} D. N. Aristov, Phys. Rev. B. {\bf 55}, 8064 (1997) --
1D+2D+3D-systems.
\bibitem{5} If carriers are
delivered by the magnetic impurities placed in the points of the
face-centered cubic lattice, their Fermi momentum equals
$k_{F0}a=2\pi(3pa^3/16\pi)^{1/3}$. For the diluted magnetic
semiconductor Ga$_{1-x}$Mn$_x$As, $k_{F0}a\approx 4.91(\gamma
x)^{1/3}$ where $\gamma<1$ is the fraction of those Mn-ions that
substitute Ga-atoms and turn into acceptors. Herefrom, it follows,
for instance, $k_{F0}=1,\,1.5,\,2$ and $2.5$ for $\gamma
x\approx0.01,\,0.03,\,0.07$ and $0.13$, correspondingly ($p=4\gamma
x/a^3=2.16\cdot10^{22}\gamma x$ cm$^{-3}$).
\bibitem{6} Another way to express $T_{\rm C}$  in terms of $H_{\rm
RKKY}$ is to relate it to the minimum effective field (at the sphere
surface): $k_BT_{\rm C}\sim \mu H^S_{\rm RKKY}$.  The Curie
temperature so defined is half as large as that shown in Fig. 2, 3.
\bibitem{7}  H. Raebiger, A. Ayuela and J. von Boehm,
    Phys. Rev. B {\bf 72}, 014465 (2005)
\bibitem{8} D. J. Priour Jr. and S. Das Sarma,
    Phys. Rev. B {\bf 73}, 165203 (2006).

\end{thebibliography}
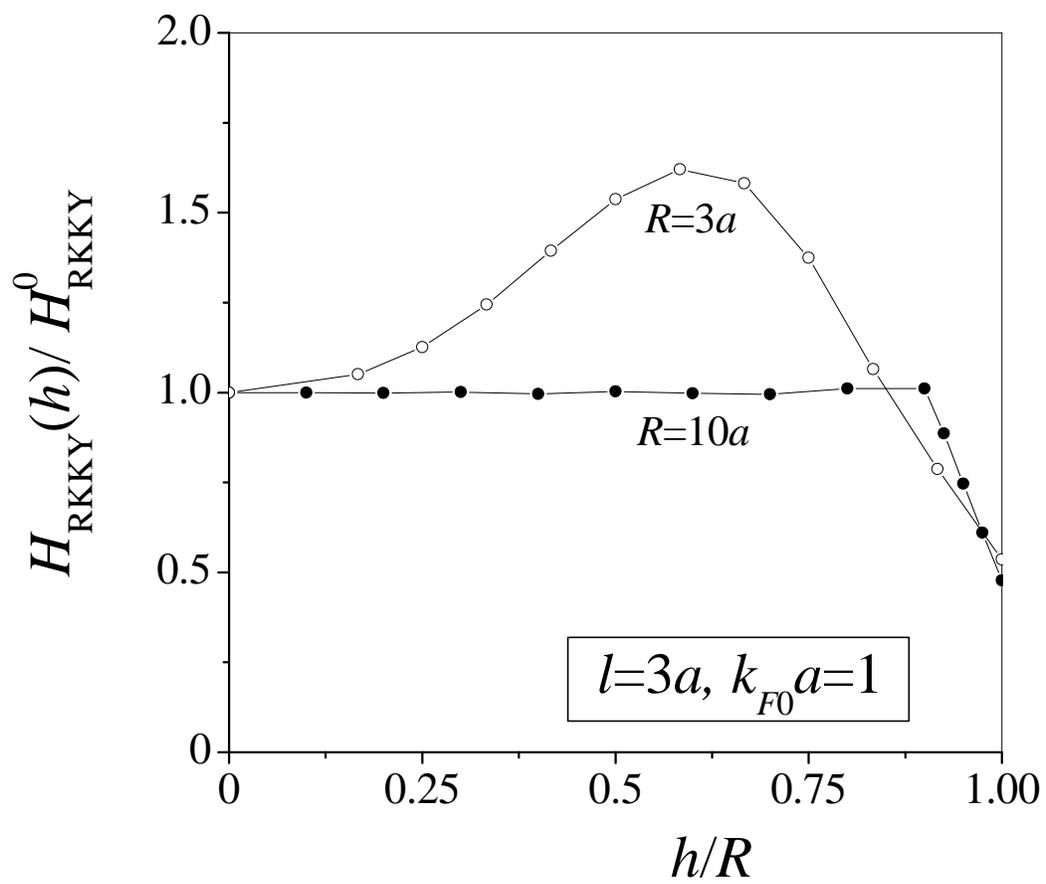
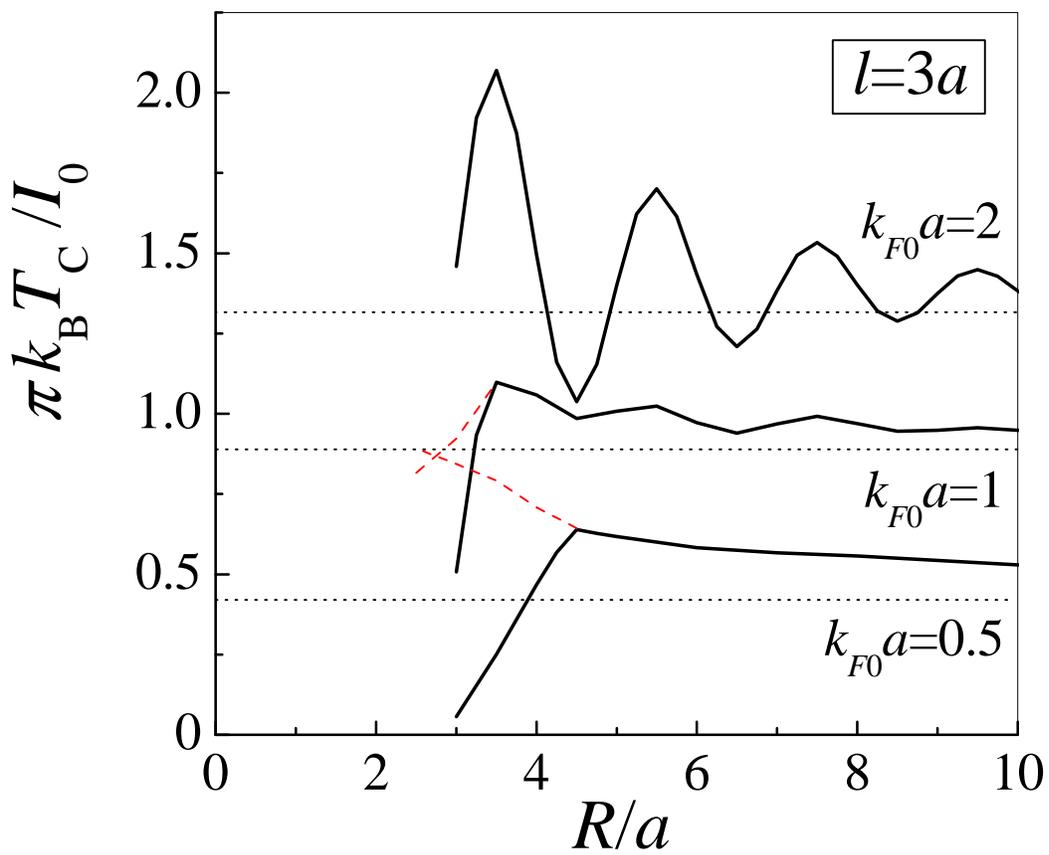
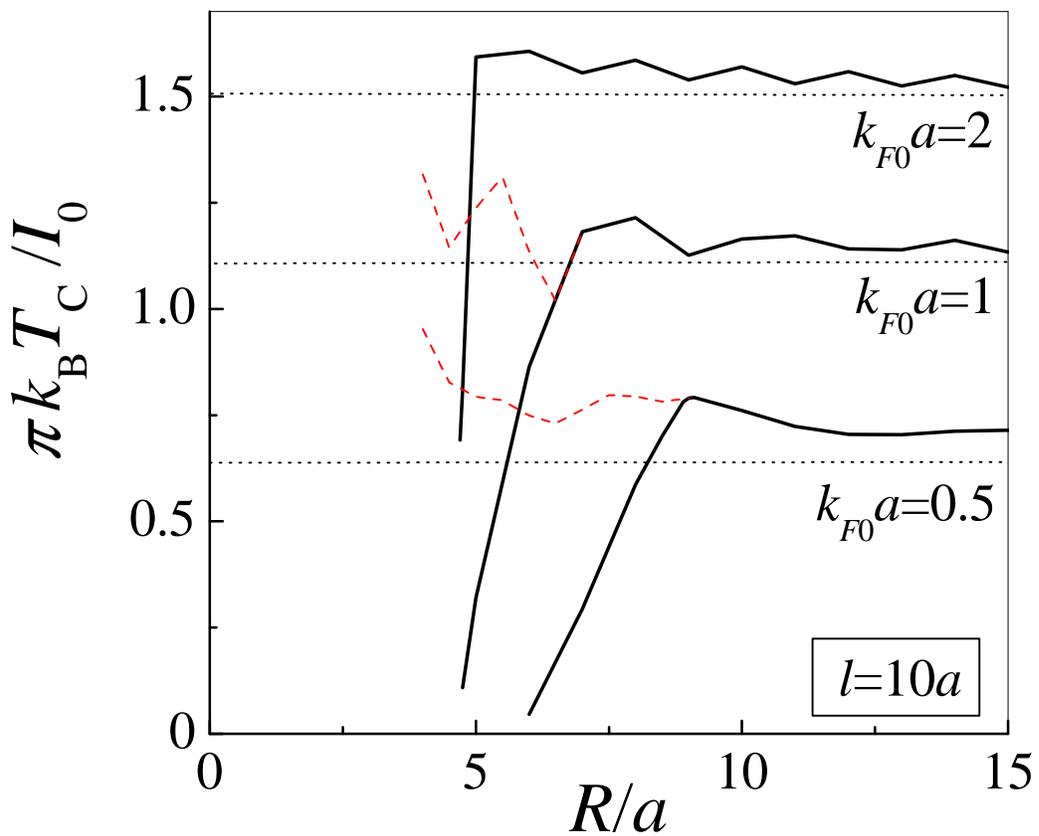
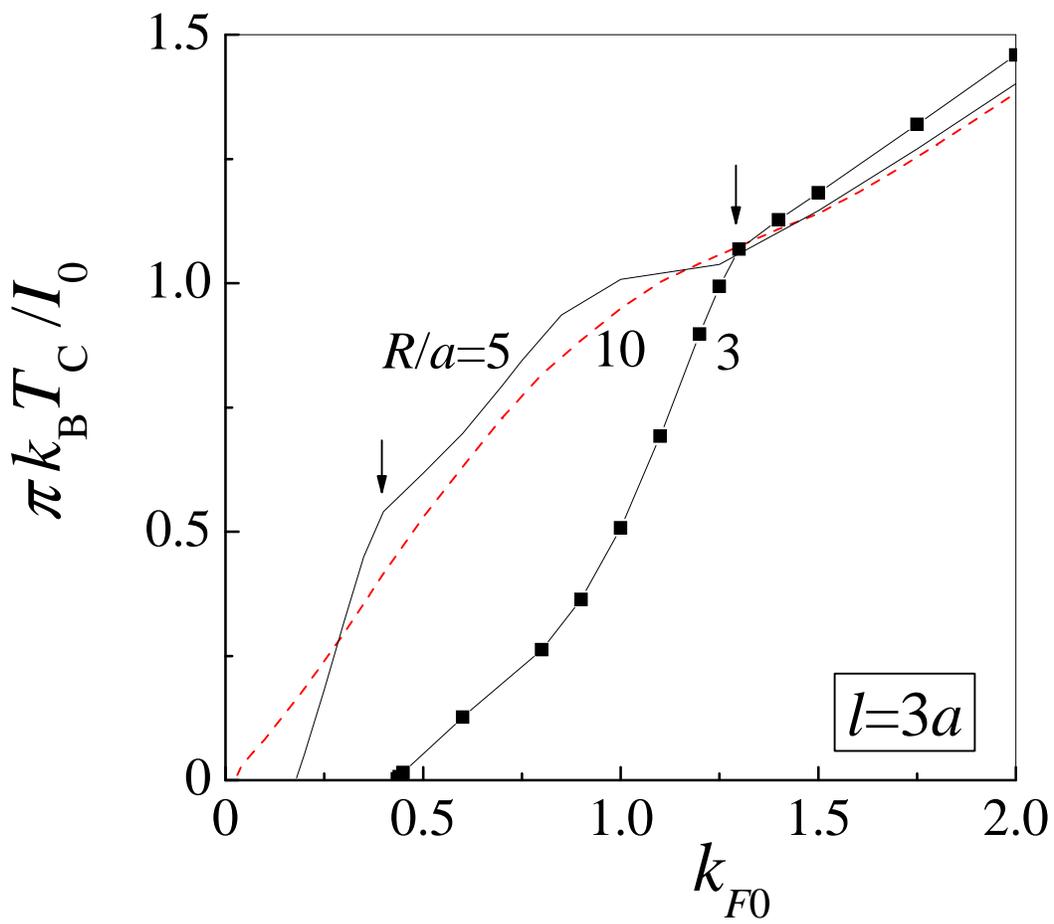
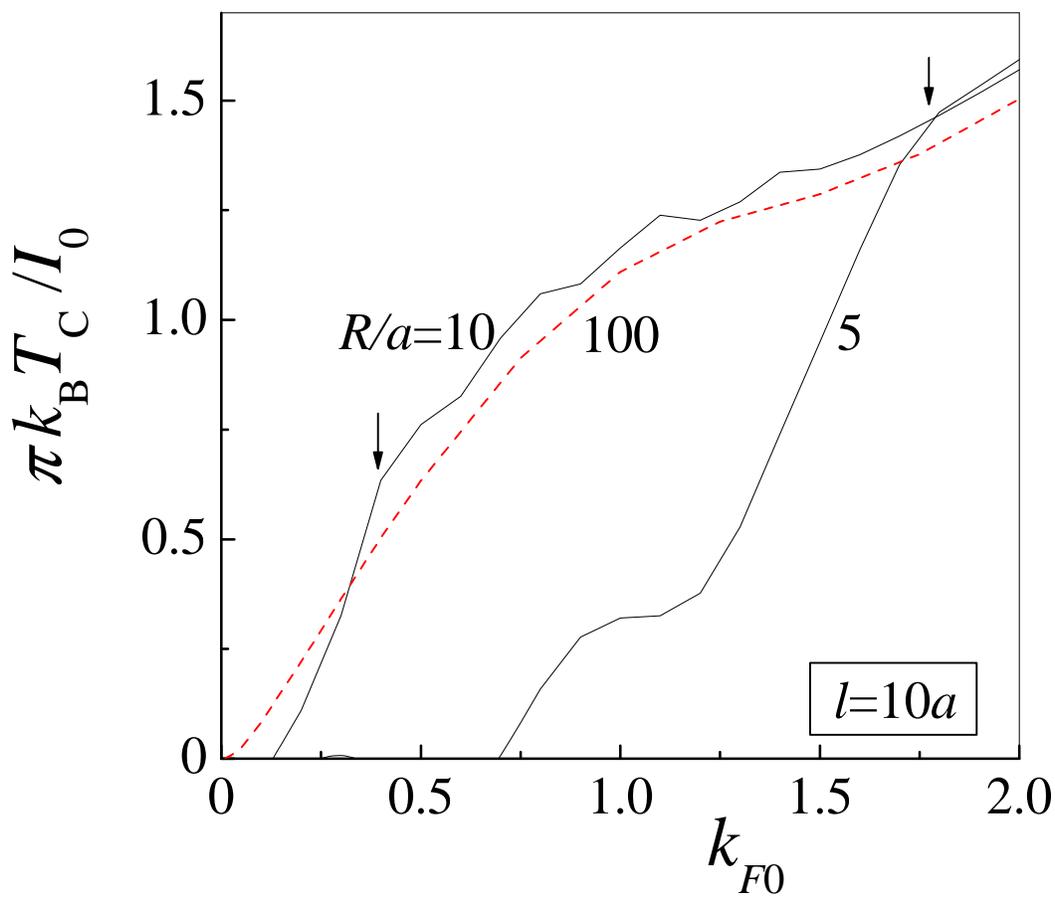
\end{document}